\documentclass{aa}  
\usepackage{graphicx}
\usepackage[normalem]{ulem}
\usepackage{txfonts}
\usepackage{hyperref}
\usepackage{graphicx}	
\usepackage{amsmath}	
\usepackage{hyperref}
\usepackage{gensymb}
\usepackage{url}
\usepackage{multirow}
\usepackage{float}
\newcommand{\ero}{\textit{eROSITA}}
\newcommand{\msun}{\(h^{-1}\,M_\odot\)}

\newcommand{\ergpspcm}{erg \,s$^{-1}$\,cm$^{-2}$}

\usepackage{silence}
\begin{document} 
   \title{Forecasts for cosmological measurements based on the angular power spectra of AGN and clusters of galaxies in the SRG/\ero\, all-sky survey}

   \author{S.Bykov
          \inst{1, 2}\thanks{sergei.d.bykov@gmail.com},
          M.Gilfanov\inst{1,3}
          \and
          R.Sunyaev\inst{1,3}\fnmsep
          }

   \institute{Max Planck Institute for Astrophysics, Karl-Schwarzschild-Str 1, Garching b. München D-85741, Germany
         \and
Kazan Federal University, Department of Astronomy and Satellite Geodesy, 420008 Kazan, Russia
             \and
             Space Research Institute, Russian Academy of Sciences, Profsoyuznaya 84/32, 117997 Moscow, Russia
             }

   \date{Received XX; accepted XX}

 
  \abstract
   {The \ero\, X-ray telescope aboard the SRG orbital observatory, in the course of its all-sky survey, is expected to detect about three million active galactic nuclei (AGN) and $\sim$ hundred thousand clusters and groups of galaxies. Such a sample, clean and uniform, complemented with redshift information, will open a new window into the studies of the Large-Scale structure (LSS) of the Universe and the determination of its cosmological parameters.} 
   {The purpose of this work is to assess the prospects of cosmological measurements with  the \ero\, sample of AGN and clusters of galaxies. We assume the availability of photometric redshift measurements for \ero\, sources and explore the  impact of their quality on our forecasts.}
   {As the LSS probe, we use the redshift-resolved angular power spectrum of the density fluctuations of objects. We employ a Fisher-matrix formalism and assume flat $\Lambda$CDM cosmology to forecast the constraining power of \ero\, samples of AGN and clusters of galaxies. We compute the LSS-relevant characteristics of AGN and clusters in the framework of the halo model and their X-ray luminosity functions. As the baseline scenario, we consider the  full 4 years long all-sky survey and investigate the impact of reducing the survey length to 2 years.}
   {We find that the accuracy of photometric redshift estimates has a more profound effect on  cosmological measurements than the fraction of catastrophic errors. Under realistic assumptions about the photometric redshift quality, the marginalized errors on the cosmological parameters achieve $1-10\%$ accuracy depending on the cosmological priors used from other experiments. The statistical significance of BAO detection in angular power spectra of AGN and clusters of galaxies considered individually achieves  $5-6\sigma$. Our results demonstrate that the \ero\, sample of AGN and clusters of galaxies  used in combination with currently available photometric redshift estimates will provide cosmological constraints on a par with  dedicated optical LSS surveys. 
   }
   {}

   \keywords{X-rays: galaxies; X-rays: galaxies: clusters; Cosmology: large-scale structure of Universe} 
\titlerunning{\ero\, cosmological forecast - Angular clustering of AGN and Clusters}
\authorrunning{Bykov, Gilfanov \&  Sunyaev}
   \maketitle

\section{Introduction}
\label{intro}
The distribution of matter in the Universe is not uniform. Galaxies form the so-called cosmic web: filaments, walls, and voids, which comprise the large-scale structure (LSS) of the Universe \citep{Dodelson2003, Eisenstein2005,  Beutler2011, Alam2017}. Mapping  the LSS yields a wealth of important information about  cosmological parameters, the evolution of density perturbations, the growth of structures, and the mass-energy content of the universe \citep{Tegmark2004, Percival2010, Blake2011, Padmanabhan2012, DESCollaboration2021a, DESCollaboration2021b}.  The LSS is traced not only by normal galaxies but also by Active Galactic Nuclei (AGN) and clusters of galaxies.

The large-scale structure has been probed by several wide-angle optical galaxy surveys such as SDSS (Sloan Digital Sky Survey)\footnote{\url{http://sdss.org}}, DES (Dark Energy Survey\footnote{\url{https://www.darkenergysurvey.org}}), 2dF (2dF Galaxy Redshift Survey\footnote{\url{http://www.2dfgrs.net}}). Using spectroscopic or photometric  redshifts of the objects,  such surveys unveil the 3-D distribution  of visible matter in the Universe, permitting to measure its power spectrum and to detect and quantify various cosmologically significant effects such as Baryon Acoustic Oscillations \citep[BAO, ][]{Sunyaev1970, Peebles1970} and redshift-space distortions \citep[RSD, ][]{Kaiser1987}. This data allows one to measure cosmological parameters \citep{Cole2005, Eisenstein2005, DESCollaboration2021a, DESCollaboration2021b, Alam2021}.  Planned space missions and ground-based facilities, such as Euclid \footnote{\url{https://sci.esa.int/web/euclid}}, Vera Rubin observatory\footnote{\url{https://www.lsst.org}}, Dark Energy Spectroscopic Instrument\footnote{\url{https://www.desi.lbl.gov}}, and Nany Grace Roman Telescope\footnote{\url{https://roman.gsfc.nasa.gov}} will  dramatically increase the number of objects available for cosmological studies, potentially increasing the accuracy of cosmological measurements.  

To date,  all cosmologically significant large-scale structure surveys have been conducted in visible light. However, the selection of galaxies and quasars in optical wavelengths is complicated  by a number of effects, such as contamination by stars and nearby galaxies,  absorption,  and several others. On the contrary, X-ray surveys provide an efficient tool for identifying accreting supermassive black holes -- indeed, AGN constitute  the majority of sources detected in extragalactic X-ray surveys \citep[e.g.][]{Brandt2005, Sunyaev2021}. 
Deep (but relatively narrow, up to $\sim 25$ sq. degrees) X-ray surveys performed by XMM-Newton and Chandra observatories demonstrated the feasibility and usefulness of  studying  clustering of X-ray-selected AGN \citep{Allevato2011, Mountrichas2016, Kolodzig2017, Kolodzig2018, Allevato2019}.  
Furthermore, hot intracluster medium shines in X-rays by which means one can detect a cluster of galaxies without actually registering individual galaxies. X-ray data allows the determination of cluster mass and temperature, quantities essential for astrophysics and cosmology \citep[e.g.][]{Vikhlinin2009, Vikhlinin2009b, Allen2011}.
Thus, X-ray-selected samples present a reasonably clean and  complete flux-limited census of extragalactic objects -- AGN and clusters of galaxies, which number density is sufficient for studying the LSS.

The SRG orbital X-ray observatory \citep{Sunyaev2021} was launched to the halo orbit around the Sun-Earth Lagrangian L2 point on July 13, 2019, and  on December 12, 2019, started an all-sky survey, which was planned to continue  for 4 years, until December 2023. So far, \ero\, telescope completed 4.4 full sky surveys and currently is in safe mode. The SRG observatory continues science observations in the interests of Mikhail Pavlinsky ART-XC telescope  -- the second science instrument aboard the SRG observatory.   

In the course of the full four-year long all-sky survey, the \ero\, telescope \citep{Predehl2021} aboard  SRG is expected to detect in the $0.5-2$ keV energy range about $\sim3$ million   AGN with a median redshift of $z\sim1$ \citep{Kolodzig2013a} and $\sim 10^5$  clusters of galaxies with a median  redshift of $z\sim0.4$  \citep{Merloni2012, Pillepich2012}. Cosmological measurements and the nature of dark energy  is the main scientific driver of the mission \citep{Sunyaev2021}. To this end, a significant role will be played by the measurements  of the mass function of clusters of galaxies based on their X-ray properties. Cosmological studies can be also conducted using the 3-D distribution of quasars and clusters of galaxies.  The enormous size  of the X-ray-selected AGN sample makes their study  especially meaningful in this context -- as it was shown earlier, BAOs will be clearly detectable in the \ero\, AGN sample \citep{Hutsi2014b, Kolodzig2013b}.  Prospects of cosmological studies with clusters of galaxies have been previously investigated by \cite{Pillepich2012}, however, the   potential role of AGN remained so far unexplored.

The purpose of the present paper is to fill this gap and to forecast the accuracy of  cosmological measurements based on samples of AGN detected in the SRG/\ero\, all-sky survey. To facilitate the comparison with AGN and with  previous works and to investigate the power of combined AGN-cluster estimates we also present  results of similar calculations  for clusters of galaxies.   In our calculations we use the characteristics of \ero\, performance measured in-flight during the first two years of the survey and realistic, currently achieved parameters of the photometric redshift estimates of AGN and clusters of galaxies.  

The paper is structured as follows. We assess the usefulness of AGN and clusters for large-scale structure studies in the presence of photometric redshift errors in section \ref{initial_feeasibility}. In section \ref{modelling} we describe the calculations of redshift distributions and linear bias factors of both tracers, as well as the model for photo-z scatter in distance measurements. We explain formalism for computing the two-point correlation function and Fisher matrices forecasts in sect. \ref{aps_and_fisher}. Sect. \ref{results} presents our forecast for BAO detectability and cosmological precision.  We discuss our findings and place the results in the context of current cosmological probes in sect. \ref{discusion}. We conclude in  section \ref{conclusions}. In appendix \ref{mcmc} we justify the usage of the Fisher matrix method by comparing its predictions with the posterior distribution calculated with the  Markov chain Monte Carlo algorithm.  

We use decimal logarithms throughout the paper and assume fiducial cosmological parameters $H_0=70$ km s$^{-1}$ Mpc$^{-1}$, $\Omega_m=0.3$, $\Omega_b=0.05$, $\sigma_8=0.8$ and $n_s=0.96$ for flat $\Lambda$CDM cosmology. Mass is in units of $M_{500c}$. Cosmological calculations of distances, halo model, biases, etc are done in Core Cosmology Library \citep[CCL\footnote{\url{https://ccl.readthedocs.io/en/latest/}},][]{Chisari2019}. Unless stated otherwise, angular power spectra are calculated with Code for Anisotropies in the Microwave Background \citep[CAMB\footnote{\url{https://camb.readthedocs.io/en/latest/index.html}},][]{Lewis2011}.

\section{Initial feasibility study}
\label{initial_feeasibility}
The three-dimensional distribution of matter is  usually  analysed with two-point statistics. A common choice is the power spectrum which shows the amplitude of density fluctuation on a given co-moving scale, and its errors are directly related to the ability of the survey to sample the LSS.
\citet{Hutsi2014b} showed that the density of AGN in  the \ero\, survey would provide sufficient signal-to-noise at the median redshift of $z=1$.

The capability of a survey to probe LSS depends on the volume of the universe observed given the power spectrum of tracer objects and their redshift distribution. This is quantified by the  effective volume of a survey, $V_{\rm eff}$ 
\citep{Eisenstein2005}:
$$V_{\rm eff} (k) = \Omega \int_{z_{\rm min}}^{z_{\rm max}}\displaylimits \left( \frac{\mathcal{N}(z) P_{\rm tr}(k,z)}{\mathcal{N}(z) P_{\rm tr}(k,z)+1} \right)^2 \frac{dV(z)}{dz} dz$$

where $\Omega$ is a solid angle of a survey, $\mathcal{N}(z)$  is the spatial number density, $P_{\rm tr}(k)$ is the tracer power spectrum, and $\frac{dV(z)}{dz}$ is the differential co-moving volume. Differential volume is calculated with the formula $\frac{dV(z)}{dz} = c r(z)^2 / H(z)$ [Mpc$^3$ sr$^{-1}$], where $r$ is the co-moving radial distance, $H(z)$ is a Hubble parameter, $c$ is the speed of light. The details of the calculation of redshift distribution and power spectrum may be found in the next section. The error on the power spectrum is proportional to the $V_{\rm eff}^{-\frac{1}{2}}$ assuming gaussian statistics and independent  bins in $k$-space.
Effective volume for the spectroscopic \ero\, sample of AGN was calculated in \citet{Kolodzig2013b} and shown to be larger than that for some of the optical surveys. In the present paper, we make one step further and include clusters of galaxies and the effects of photometric redshift errors.

We utilize the model of \citet{Hutsi2010} to take into account the power suppression on small scales due to the photo-z errors. The power spectrum of tracers is found as 
$$P(k, z)_{\rm tr} = b^2(z) P(k, z)_{\rm CDM} \times \frac{\sqrt{\pi}}{2\sigma k} \operatorname{erf}(\sigma k )$$
where $b(z)$ is the linear bias factor, $ P(k, z)_{\rm CDM}$ is the power spectrum of dark matter, and spatial suppression  scale $\sigma=\frac{c}{H(z)}\sigma_z$ ($c$ is the speed of light, $H(z)$ is the Hubble constant at given $z$, and $\sigma_z = \sigma_0(1+z_{\rm eff})$ is the photometric redshift error). $z_{\rm eff}$ is taken as $1$ for AGN and $0.4$ for clusters. $\sigma_0$ is a free parameter which depends on the quality of the optical follow-up. We restrict $0.5<z<2.5$ and $0.1<z<0.8$ for AGN and clusters respectively, assuming 65\% of the sky used in the analysis. Photo-z scatter $\sigma_0$ is varied between 0 and 0.05. 

\begin{figure}
    \centering
    \includegraphics[width=\hsize]{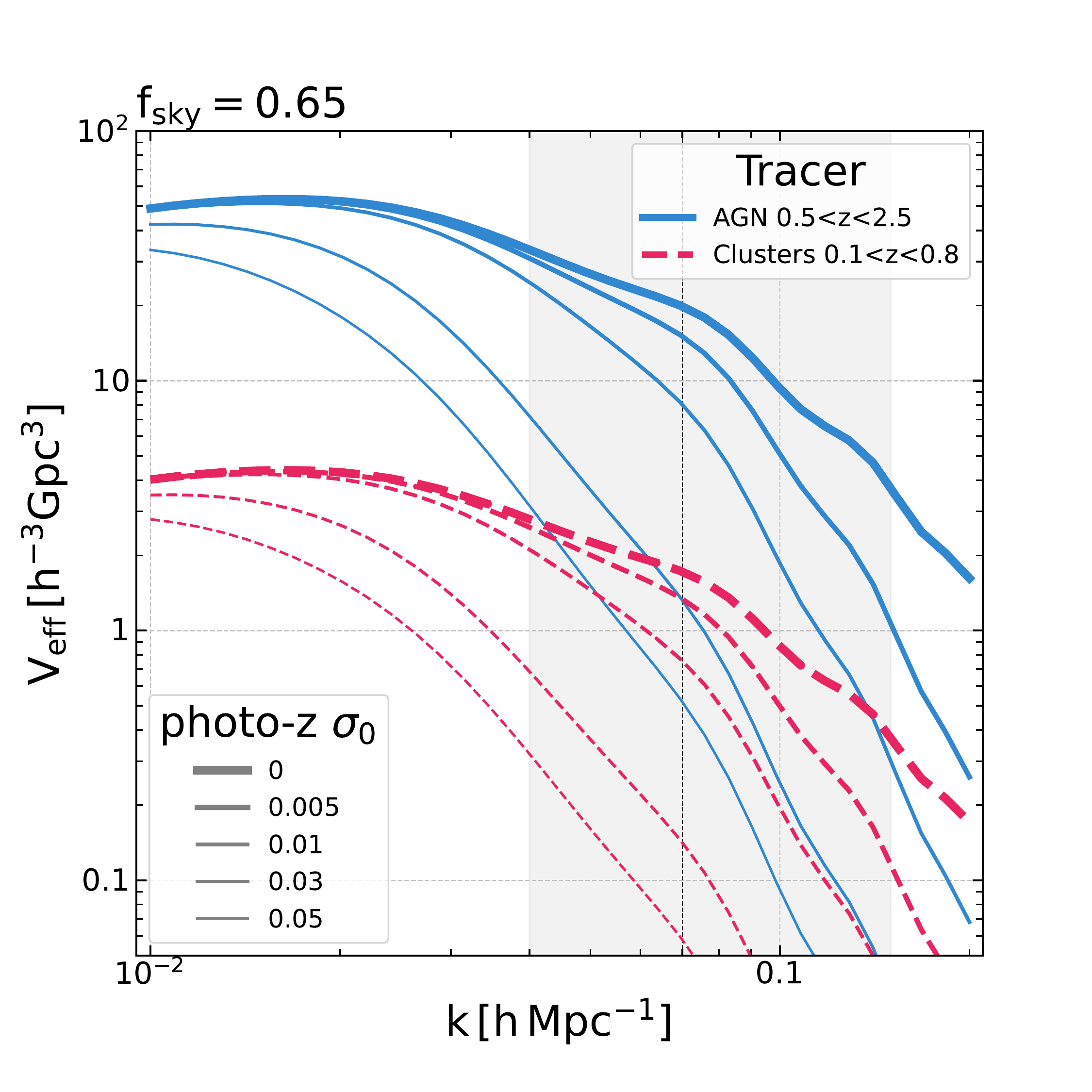}
    \caption{Effective volume as a function of co-moving scale probed by AGN and clusters (solid and dashed lines respectively) in the \ero\, all-sky survey. The thicker the line, the better the quality of photo-z. The grey band illustrates the scales where the BAO feature is prominent, and the vertical dashed line shows the position of the first peak.} 
    \label{fig:veff}
\end{figure}

We show the results in Fig. \ref{fig:veff}. The effective volume for AGN sample drops from $\sim50$  to less than $\sim2$ h$^{-3}$Gpc$^3$ between  $k=0.01$ and $k=0.2$ h Mpc$^{-1}$. For clusters the figures are $\sim4$  and $\sim0.2$ h$^{-3}$Gpc$^3$.
 It can be seen that on very large scales (small $k$) the AGN sample is more efficient than clusters even when the photo-z errors are large. 
 In the area of moderate scales ($k=0.05-0.2$) the relative performance of both tracers depends significantly on the quality of photo-z. For instance, even though AGN are much more numerous (2 mils), with the $\sigma_0=0.05$ they would be outperformed by a much smaller number of clusters (90k) with $\sigma_0<0.005$. Typically one may expect $\sigma_0=0.03$ for AGN and $0.01$ for clusters. In this configuration, the effective volumes at BAO scales are comparable. 
 
We, therefore, expect that both tracers suffer significantly from the loss of the photometric redshift precision and that the AGN and cluster samples would provide roughly comparable results and  complement each other.
 
\section{Modelling details}
\label{modelling}
In this section, we illustrate our calculations of redshift distributions and number counts (sect. \ref{xlf}), linear bias factors (sect. \ref{bias}) and the model for photometric redshift (sect. \ref{photoz}). 
\subsection{X-ray luminosity functions}
\label{xlf}
To the power spectrum shape, we need to know the space number densities of AGN and clusters.

With the knowledge of the X-ray luminosity function (XLF) of AGN $\phi(logL,z)$ and the flux limit of the survey, we calculate the  distribution of AGN with redshift. We use the XLF measured in \citet{Hasinger2005} with an exponential cutoff for $z>2.7$ \citep{Brusa2009}. For details of calculations see \citet{Kolodzig2013a}. 

The shape of the AGN XLF is the following:

$$
\phi(L,z) = K_0 \left[ \left( \frac{L}{L_*} \right)^{\gamma_1} + \left( \frac{L}{L_*} \right)^{\gamma_2}  \right]^{-1}\times e(L,z)
$$\
where factor 
\begin{equation}
    e(L,z) = \begin{cases}
     (1+z)^{p_1}, z\leq z_c(L)\\
      (1+z_c(L))^{p_1} \left( \frac{1+z}{1+z_c(L)} \right)^{p_2}, z > z_c(L)
    \end{cases}
\end{equation}
and cutoff redshift 
\begin{equation}
    z_c(L) = \begin{cases}
     z_{c,0} \left( \frac{L}{L_\alpha} \right)^\alpha, L\leq L_\alpha \\
      z_{c,0}, L>L_\alpha
    \end{cases}
\end{equation}

All parameters were taken from \citet{Hasinger2005}, table 5, LDDE model. Cutoff over redshift 2.7 is done as in \citet{Kolodzig2013a}, i.e. $\phi(L,z) \rightarrow \phi(L,2.7)\times10^{0.43(2.7-z)}$ for $z>2.7$. 

The redshift distribution $n(z)$, per deg$^2$, is found as
$$n(z) = \frac{dV(z)}{dz} \int_{log L_{\rm min}(S,z)}^{48}\displaylimits \phi(logL,z) dlogL$$
 where $\frac{dV(z)}{dz}$ is the differential co-moving volume, $S$ is the flux limit of the survey (we adopt $10^{-14}$ \ergpspcm), and the smallest observed luminosity $L_{\rm min}$ at given redshift $z$ is found as $L_{\rm min} = 4\pi S r^2_{\rm L}$ ($r_{\rm L}$ is the luminosity distance). The k-correction was done assuming photon index $\Gamma=1.9$ and no absorption.
We multiply the XLF by a factor of 1.3 to match the predicted sky number density of AGN at flux limit $10^{-14}$ \ergpspcm\, with observations (i.e. $\sim90$ AGN/deg$^2$), see, e.g.,  \citet{Georgakakis2008} 

The resulting distribution of AGN is shown in Fig. \ref{fig:dndz} (panel A). The distribution peaks at $z\approx1$ and yields $\approx 2474000$ AGN on 65.8\% of the sky. This sky fraction corresponds to the extragalactic sky  $|b|>15$ deg. During the calculations of the effective volume in the previous section, the spatial number density  of objects was found as $\mathcal{N}(z)=\frac{n(z)}{dV/dz}$.

To calculate the distribution of clusters of galaxies we invoke the halo model approach. We start from the halo mass function $n(M, z)$ of \citet{Tinker2008}  and  the fixed mass-luminosity relation of X-ray clusters calibrated in \citet[eq. 22]{Vikhlinin2009}\footnote{We ignore the scatter in this relation for simplicity, but uncertainties in this relation have a profound effect on the cosmological analysis, see \citet{Pillepich2012}.}. We use k-correction assuming thermal plasma model $apec$ with temperature from  \citet[table 3, free slope]{Vikhlinin2009}. The minimum mass for integration is taken as $5\times 10^{13}$ \msun\,\footnote{$5\times 10^{13}$ \msun = $7.14\times 10^{13} M_{\odot}$ in our fiducial cosmology} since the ML relation is not well-known for lower masses.
The redshift distribution is then found as:
$$n(z) = \frac{dV(z)}{dz} \int_{M_{\rm min}(S,z)}^{M_{\rm max}}\displaylimits n(M,z) dM$$
where, as before, $M_{\rm min}(S,z)$ is the minimum observable  mass at a given redshift, and $M_{\rm max}$ is the maximum mass considered ($10^{16}$ \msun). As the limiting flux we assume $4.4\times 10^{-14}$ \ergpspcm\, for extended source detection, see \citet[fig. 4]{Pillepich2012}. 

Such flux limit produces redshift distribution shown in panel B in Fig. \ref{fig:dndz}, with the break at $z\approx0.3$ due to the minimum mass cut. On 65.8\% of the sky, one would detect $\approx$93400 clusters, with $\approx$ 900 of them at $z>1$. Approximately 52200 would have mass $M>10^{14}$ \msun\,, and 3100 clusters with  $M>3\times 10^{14}$ \msun. 

\subsection{Linear bias factors}
AGN and clusters live in dark matter halos.  Halos are biased tracers of the underlying dark matter distribution, and they are more clustered. 

The bias factors of X-ray-selected AGN at different redshifts were measured with XMM-Newton data \citep{Allevato2011} and were found to be consistent with the biasing of dark matter halos with the mass similar to that of the galaxy group, $M_{\rm DMH} \sim 10^{13}$ \msun. We, therefore, use the model (fitting formulae) of \citet{Tinker2010} for bias $b_{\rm eff}(z)$ of dark matter halos with mass of $2\times10^{13}$  \msun. The bias is shown in panel C of Fig. \ref{fig:dndz}. 

For X-ray clusters we use the halo model with effective bias given by the average bias of halos weighed with their number density:

$$b_{\rm eff}(z) = \frac{\int_{M_{\rm min}(S,z)}^{M_{\rm max}} b(M,z) n(M,z) dM}{n(z)}$$
We show the effective bias of the cluster population in Fig. \ref{fig:dndz}, panel C. At the peak redshift of the sample, the bias factor is $\sim3$. This is consistent with the bias measurements with the two-point correlations functions of $\sim 200$ X-ray-selected galaxy clusters from the XMM-Newton XXL survey \citep{Marulli2018}. We limit our calculations for clusters of galaxies by a redshift of 0.8, as explained in Section 4. Correspondingly, we do need to construct a bias model for clusters beyond this redshift value.
\label{bias}

\subsection{Photometric redshift model}
\label{photoz}

Photometric redshifts have poorer accuracy than redshifts based on spectroscopic data. However, it is compensated by the possibility to estimate redshifts for a much larger number of objects.
A common approach in analyzing photometric-redshift data sets is to bin objects in the  redshift space so that one 'integrates out' the effects of photo-z errors. The width of the bins is chosen to be a few times the photo-z scatter at a given z. The true underlying distribution of redshifts (spec-z) would be smoothed in those bins by random errors (i.e. scatter $z_{\rm spec} - z_{\rm phot}$). We use the model of \citet{Hutsi2014b} for gaussian photometric redshift errors (see their eq. 7) with two parameters: $\sigma_0$ which controls the scatter and $f_{\rm fail}$ - the catastrophic errors fraction. The smoothed shape of $i$-th redshift bin $n^{(i)}$ ($z_{p_1}^{(i)}<z<z_{p_2}^{(i)}$) is given by

$$
\begin{aligned}
n^{(i)}(z)=n(z) &\left[\left(1-f_{\text {fail }}\right) \frac{\operatorname{erf}\left(\frac{z_{\mathrm{p}_{2}}^{(\mathrm{i})}-z}{\sqrt{2} \sigma(z)}\right)-\operatorname{erf}\left(\frac{z_{\mathrm{p}_{1}}^{(\mathrm{i})}-z}{\sqrt{2} \sigma(z)}\right)}{1+\operatorname{erf}\left(\frac{z}{\sqrt{2} \sigma(z)}\right)}\right.\\
&\left.+f_{\text {fail }} \frac{z_{p_{2}}^{(i)}-z_{p_{1}}^{(i)}}{z_{p}^{\max }}\right],
\end{aligned}
$$
where $\sigma(z) = \sigma_0(1+z)$ is the scatter, and $z_p^{\rm max}$ is the maximum redshift of the sample. 

\begin{figure}
    \centering
    \includegraphics[width=\hsize]{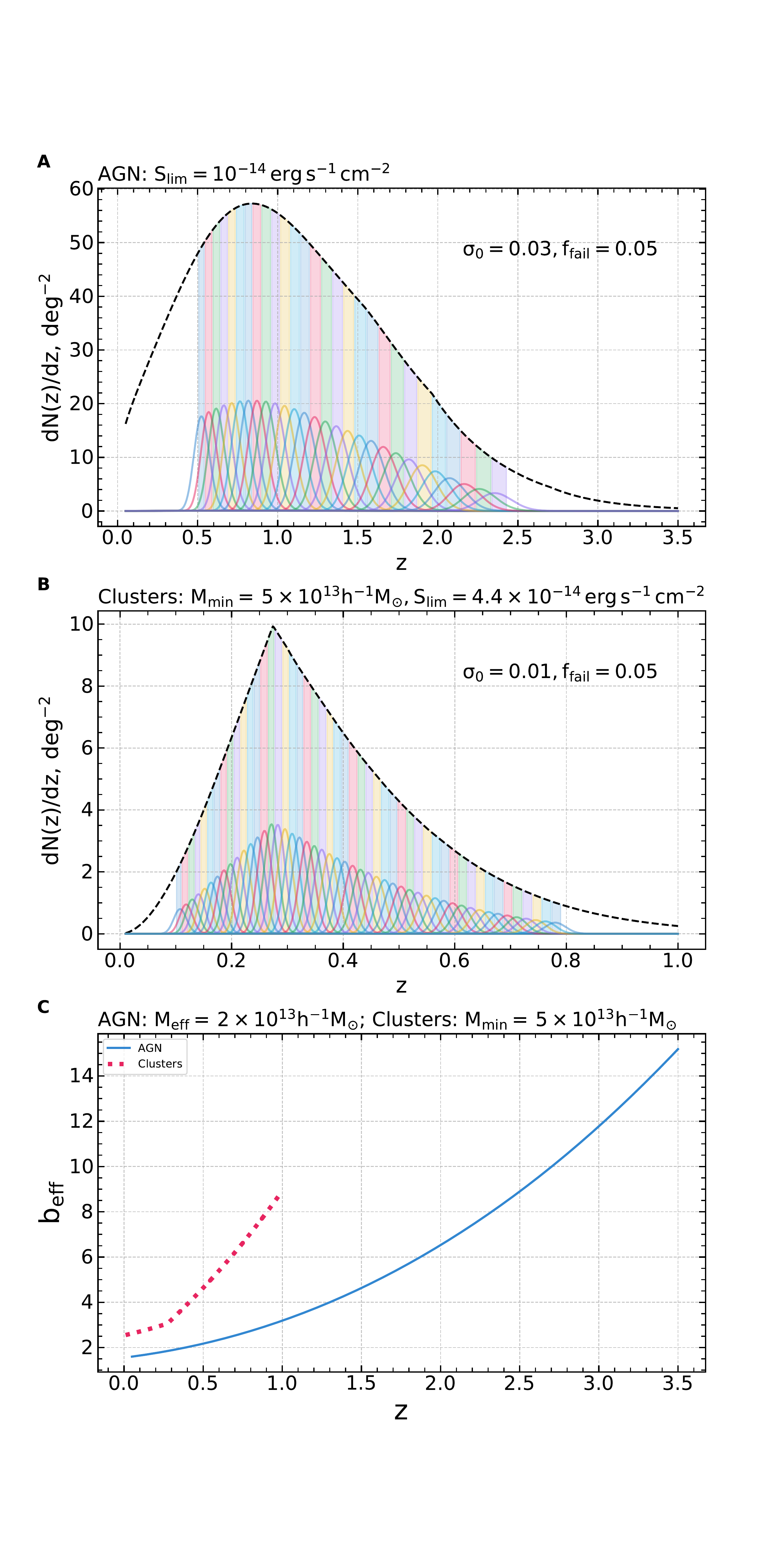}
    \caption{Panels A and B: redshift distributions of AGN and clusters tracers. The black dashed line in each panel shows the total distribution of objects, while solid lines show the distributions of objects in photo-z bins (the corresponding vertical stripes show  the boundaries of the bins in the photo-z space). The parameters for the photometric redshifts  scatter are shown in each panel. Panel C: effective linear bias factors of tracer populations as a function of $z$. }
    \label{fig:dndz}
\end{figure}

\begin{figure}
    \centering
    \includegraphics[width=\hsize]{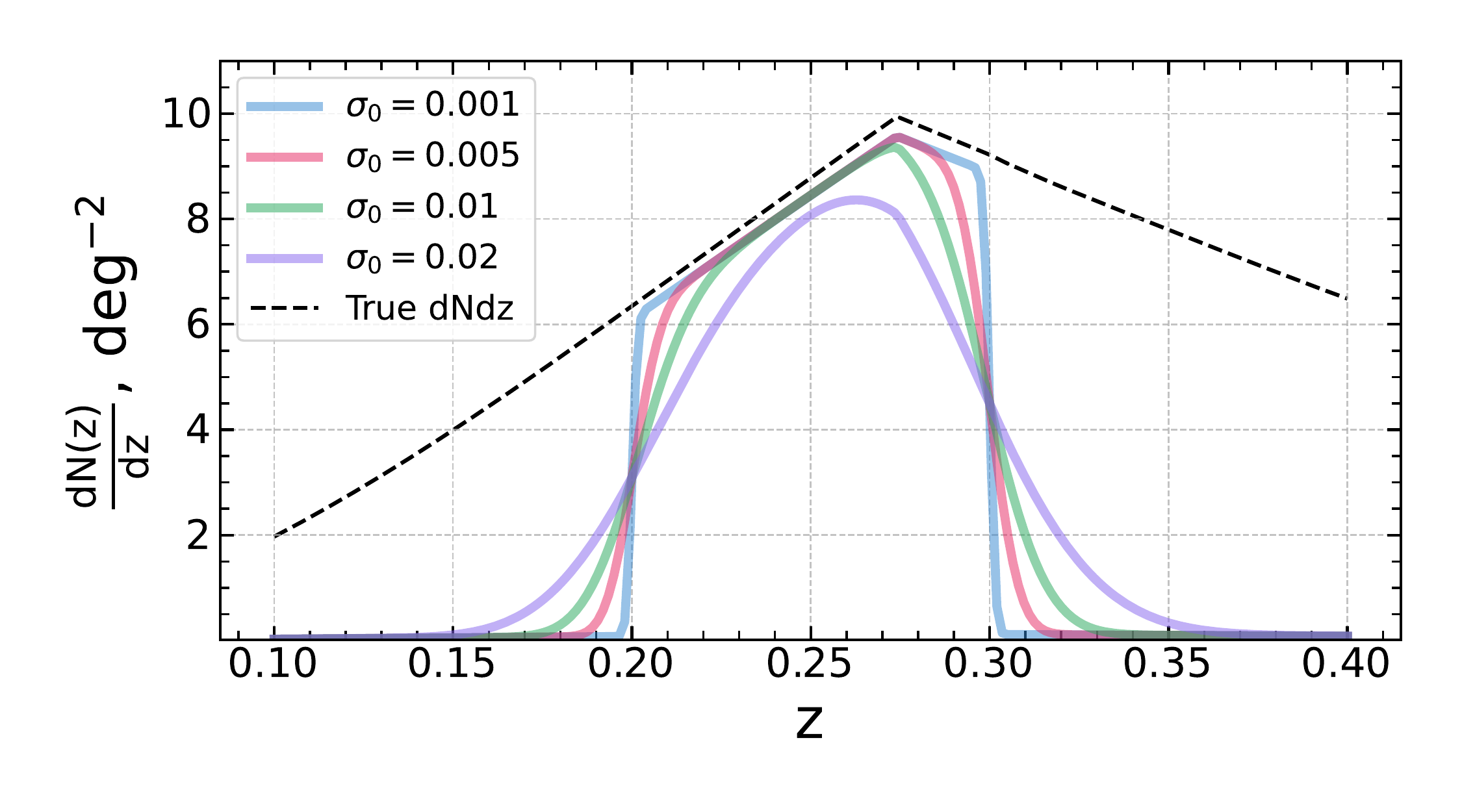}
    \caption{Effect of photometric redshift errors on the redshift distribution of clusters of galaxies.  The black dashed line shows the  true distribution of clusters of galaxies over redshift. Thick solid lines show the observed distribution of clusters from the redshift bin $0.2<z<0.3$ over photometric redshift for various values of $\sigma_0$. The values of $f_{\rm fail}$ is fixed at $0.05$. }
    \label{fig:dndz_bin}
\end{figure}

The diluted bins of photometric redshift selection of AGN or clusters for fiducial values of parameters $\sigma_0$ and $f_{\rm fail}$ are shown in Fig. \ref{fig:dndz} on panels A and B with coloured solid lines. The coloured vertical strips correspond to the bin's borders in photo-z space, and the solid line of the respective colour shows the true underlying distribution in spec-z space.  As the redshift bin width, we use $\Delta z = \sigma(z)$ at a given redshift if not stated otherwise. We illustrate the effect of photometric redshift errors in Fig.  \ref{fig:dndz_bin} where we plot the dilution of a redshift bin for various values of $\sigma_0$.

Multi-wavelength  data and the availability of rich spectroscopic information for large samples of objects allow efficient training of the  models for source classification and photo-z evaluation. The  SRGz machine learning system for classification and photometric redshift estimation is being developed by the Russian \ero\, consortium (Meshcheryakov et al., in prep., see also \citet{Borisov2021, Belvedersky2022}). SRGz is capable of determining photometric redshift probability density distribution  of individual X-ray-selected AGN with accuracy better than $\sigma_0\sim0.05$ and the outlier fraction better than $\sim10\%$.  The accuracy of photo-z for galaxy clusters is  nearly an order of magnitude  better. 
In our analysis, we assumed that photometric redshift error and fraction of catastrophic failures are known. These quantities and their redshift dependence are usually  provided by the photometric redshift estimation code, based on the analysis of verification samples of objects whose redshifts are measured in spectroscopic observations. Our approach is similar, for example to that used in \citep{Sereno2015}. In the calculations below we assumed the photo-z accuracy which is within the reach of the current version of the SRGz system.

\section{Angular power spectrum and Fisher matrices}
\label{aps_and_fisher}
In this section, we describe the computation of angular power spectra (sect. \ref{aps}) and the calculation of Fisher matrices (sect. \ref{fisher}).
\subsection{Angular power spectrum}
\label{aps}
Angular power spectrum (APS) is a Fourier counterpart of the two-point angular correlation function. The benefit of using angular correlations is that one does not assume fiducial cosmology to analyze data. 

The angular power spectrum of sources in $i$ and $j$-th redshift bins is calculated as follows:

$$\mathbf{C}_{\ell}^{ij} = \frac{2}{\pi} \int P(k, z=0) W_\ell^i(k) W^j_\ell(k) k^2dk$$
where $P(k, z=0)$ is the matter 3d power spectrum at present time and $W^i_\ell(k)$ is a projection kernel of bin $i$.
$$W^i_\ell(k)=\int j_{\ell}(kr)f^{(i)}(r)g(r)b_{\rm eff}(r)dr$$
where $r$ is a comoving radial distance, $j_\ell$ is the spherical Bessel function of order $\ell$, $f^{(i)}$ is the normalized redshift distribution of $i$-th bin, $g$ and $b_{\rm eff}$ are  linear growth and bias factors respectively. 

Calculations of angular power spectra are done in CAMB and include the effects of redshift space distortions. The matter power spectrum is calculated by CAMB in the linear regime, while bias factors and redshift distributions are found as described in section \ref{modelling}. We apply the Limber approximation \citep{Simon2007} at $\ell>110$ for both tracers. We use logarithmic bins in $\ell$. 
A minimum multipole number $\ell=10$ due to the effects of the sky mask at lower multipoles  is used in all calculations. To not include nonlinear effects in our analysis, we restrict $\ell<500~(150)$ for AGN (clusters) samples \citep{Kolodzig2013b, Hutsi2014b}. For the same reason we use 0.5 (0.1) as the minimum redshift of a cosmological sample of AGN (clusters), whilst the maximum is 2.5 (0.8) due to the scarcity of objects at higher redshifts. 

 We use the following analytical formula for the Gaussian covariance matrix of angular power spectra 
$$ \mathrm{Cov}_{\ell}(\mathbf{C}_{\ell}^{ij}\mathbf{C}_{\ell}^{mn}) = \frac{1}{(2\ell+1)f_{\rm sky}}\left( \mathbf{C}_{\ell}^{im}\mathbf{C}_{\ell}^{jn} + \mathbf{C}_{\ell}^{in}\mathbf{C}_{\ell}^{jm}\right) $$
with $f_{\rm sky}$ being the fraction of the sky observed and $\mathbf{C}^{ij}$ includes the shot noise ($1/N$, where $N$ is a surface density in the bin) if $i=j$. We treat partial sky coverage lightly with factor $f_{\rm sky}$, whereas in reality the covariance between different modes would be introduced,  especially for sky masks of complex shape. In the case of extragalactic sky $|b|>15\degr$ considered here this effect can be neglected  of this calculation. 


To accelerate computations, we ignore cross-correlation between the bins which are far-spaced (more than $3\sigma(z)$ apart). We remind the reader that the width of the redshift bin $\Delta z$ is adjusted so that $\Delta z = \sigma_0(1+z)$ for given photo-z scatter parameter $\sigma_0$ unless stated otherwise.

\subsection{Fisher matrix}
\label{fisher}

We make use of the angular power spectra  to derive errors in the determination of cosmological parameters with the Fisher matrix formalism. 

Fisher matrices approximate the posterior distribution of the parameters of interest with multidimensional Gaussian distribution \citep[][]{Tegmark1997a, Tegmark1997b, Dodelson2003}. We use the following formula for a Fisher matrix, assuming Gaussian distribution of data points and constant covariance: 

$$
F_{i,j} =\frac{\partial \mathbf{C}_{\ell}^T}{\partial \theta_i} \mathrm{Cov}_{\ell}^{-1}\frac{\partial \mathbf{C}_{\ell}}{\partial \theta_j}
$$
where $F_{i,j}$ is a Fisher matrix, $i,j$ are the indices of cosmological parameters of interest,  $\mathbf{C}_{\ell}$ is a data vector, and $\mathrm{Cov}_{\ell}$ is data covariance. 
Our model includes all five cosmological parameters of flat $\Lambda$CDM: dark matter density fraction $\Omega_{c}$, baryon density fraction $\Omega_b$, reduced Hubble constant $h$, slope of primordial power spectrum $n_s$ and the amplitude parameter $\sigma_8$. Derivatives and covariance matrices are calculated in fiducial cosmology. We transform the Fisher matrices so that instead of dark matter density $\Omega_c$ we use matter density $\Omega_m=\Omega_c+\Omega_b$, and add priors to the matrix if needed \citep{Coe2009}. 

We find derivatives numerically using {\sc numdifftools} library\footnote{\url{https://github.com/pbrod/numdifftools}} with the jacobian of data vector calculated with step $5\times10^{-4}$ and 2-point central finite difference. 

We do not include  uncertainties of the redshift distribution $n(z)$, bias $b_{\rm eff}(z)$ or M-L relation for clusters of galaxies in our simulations. We assumed that these quantities will be measured from the eROSITA all-sky survey with sufficient accuracy.  In particular, as it was demonstrated in \citet{Kolodzig2013a, Comparat2019} the AGN XLF and bias $b(z)$  will be accurately measured from the all-sky survey data. The M-L relation for clusters of galaxies also plays an important role in cosmological measurements with clusters (e.g. \citet{Pillepich2012}). To this end, it is  planned that \ero\, will measure this relation with high precision in the dedicated pointed phase after the all-sky survey ends.

After the Fisher matrix $F_{i,j}$ is calculated, its inverse can be used as the covariance matrix of the parameters, from which one obtains marginalized errors \citep{Tegmark1997b, Coe2009}.  In appendix \ref{mcmc} we compare the marginalized errors and credibility contours returned by the Fisher analysis and by the Markov chain Monte Carlo (MCMC) algorithm of posterior sampling and demonstrate that they produce consistent results.

\begin{figure*}
    \centering
    \includegraphics[width=\hsize]{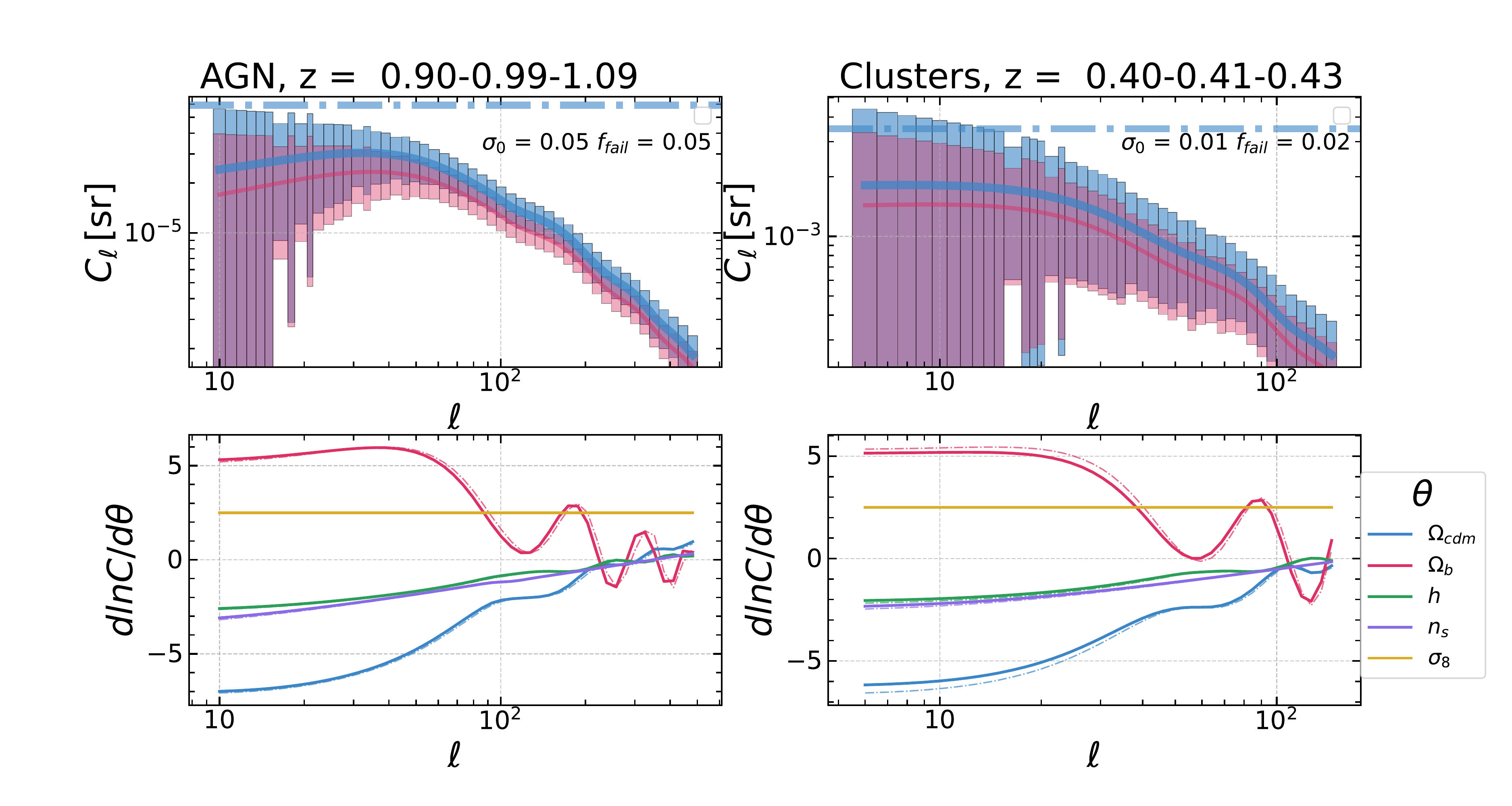}
    \caption{Example of angular power spectra of AGN (left panels) and clusters (right panels). On the top panels the auto-spectrum of the first redshift bin is shown in blue (bin 0.9<z<0.99 for AGN and 0.40<z<0.41 for clusters), and the cross spectra of the first  redshift bin with the second bin (0.99<z<1.09 for AGN and 0.41<z<0.43 for clusters) in red. Horizontal lines show the level of Poisson noise for auto-spectra. On the bottom panels, the derivatives of those power spectra (solid and dashed lines, respectively) are shown with respect to parameters in the plot's legend. Parameters of photometric redshift errors are $\sigma_0=0.05$ and $0.01$, $f_{\rm fail} = 0.05$ and $0.02$ for AGN and clusters respectively.} 
    \label{fig:aps}
\end{figure*}

In Fig. \ref{fig:aps} we show the angular power spectrum examples 
for AGN and clusters. In addition, we show logarithmic derivatives  with respect to all cosmological parameters.

\section{Forecast results}
If the following section we describe the main results of the paper: the significance of BAO detection in AGN and clusters distributions (sect. \ref{results_bao}) and the forecast on cosmological constraining power  of SRG/\ero~ samples of AGN and clusters of galaxies (sect. \ref{results_cosmo}). 

To investigate the impact of photo-z errors, we use a grid of  parameters $\sigma_0$ and $f_{\rm fail}$: $\sigma_0 = 0.005, 0.01, 0.015, 0.02, 0.03, 0.05, 0.07, 0.1, 0.2,0.3$ and $f_{\rm fail} = 0.01, 0.02, 0.05, 0.1, 0.2$ for clusters and the same for AGN except for AGN we do not do calculations for $\sigma_0=0.005$ and $0.01$.   Redshift bin sizes are $\Delta z =  \sigma_0(1+z)$ except  for the case of $\sigma_0=0.005$ when we use redshift bins  $\Delta z = 1.3\times \sigma_0(1+z)$  to reduce the size of the matrices involved. We assume the sky survey of $f_{\rm sky} = 0.658$ down to fluxes $10^{-14}$ \ergpspcm ($4.4\times 10^{-14}$ \ergpspcm) and use objects in the redshift range of $0.5<z<2.5$ ($0.1<z<0.8$) for AGN (clusters). The total cosmological sample size would be $\approx1.97$ million AGN and $\approx88000$ clusters.

\label{results}
\subsection{Baryon acoustic oscillations}
\label{results_bao}
We start from the question of whether BAO would be detectable in the distribution of AGN and clusters depending on the quality of photo-z. A similar task was done in \citet{Hutsi2014b} for the AGN population. 

For BAO detection we use the following technique: given the smooth model for the 3d matter power spectrum (template without BAO wiggles, NW superscript) we find the $\chi^2$ difference between data vectors $\mathbf{C}_{\ell}$  for a model with and without BAO (i.e. confidence in units of $\sigma$):

$$
S/N = \sqrt{ (\mathbf{C}_{\ell} - \mathbf{C}_{\ell}^{NW})^T  \mathrm{Cov}_{\ell}^{-1}  (\mathbf{C}_{\ell} - \mathbf{C}_{\ell}^{NW})}
$$
As for the calculations of BAO significance, we use angular power spectra calculated by the CCL library (in contrast to the CAMB calculations done in the remainder of the paper) and compare APS obtained with the matter power spectrum of  Eisenstein and Hu with and without wiggles \citep{Eisenstein1998} in Limber approximation.

\begin{figure*}
    \centering
    \includegraphics[width=0.9\hsize]{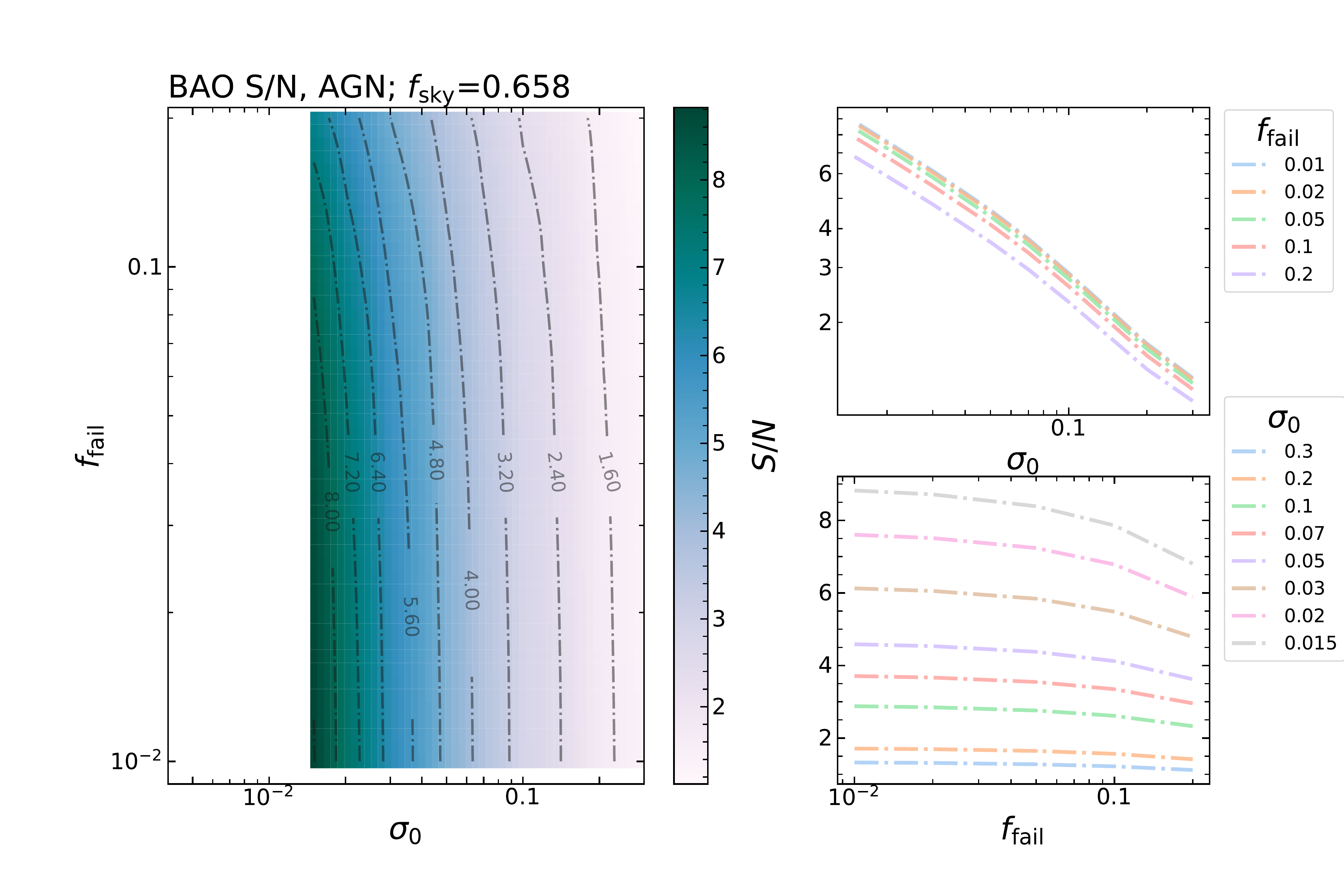}
    \includegraphics[width=0.9\hsize]{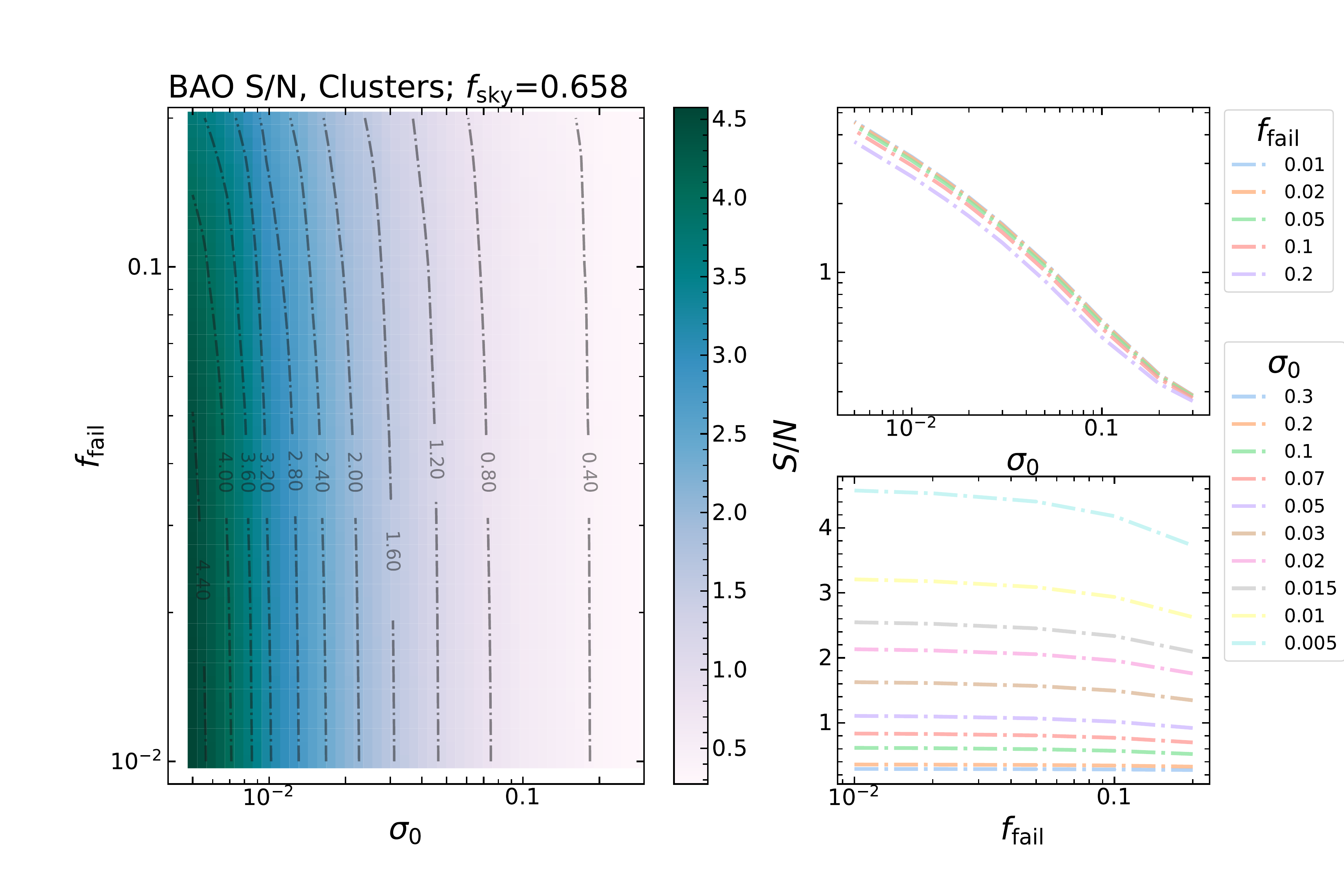}
    \caption{Significance of BAO detection (S/N) in the sample of AGN (top sub-figure) and clusters (bottom). On the left panel, the contours of significance are shown as a function of photo-z accuracy  parameters  $\sigma_0$ and $f_{\rm fail}$. On the right panels, the cross-sections with constant $f_{\rm fail}$ (top) and $\sigma_0$ (bottom) are shown.} 
    \label{fig:bao}
\end{figure*}

The results of the exercise are presented in Fig. \ref{fig:bao}. We show contours of the significance of BAO detection as a function of both $\sigma_0$ and $f_{\rm fail}$, along with cross-sections of such a surface with planes of constant  $\sigma_0$ or $f_{\rm fail}$. 

One can immediately see that in our regime for both tracers the photo-z scatter parameter has a greater impact on the performance (BAO detectability)  than the catastrophic errors fraction. Namely, the decrease of $\sigma_0$ from 0.1 to 0.03 leads to the increase in the significance of BAO detection from $\sim3\sigma$ to $\sim6\sigma$ (assuming $f_{\rm fail}=0.01$ for AGN sample). Our results are in line with the work of \citep{Hutsi2014b}, see their fig. 7, notwithstanding the difference in implementation. For AGN sample, the BAO significance achieves $\sim7-8\sigma$ if $\sigma_0=0.015$ (1.5\%), but for more realistic $\sigma=0.03$ (3\%) the figure is $\sim5-6\sigma$ with little dependence on $f_{\rm fail}$. For clusters with $\sigma_0=0.005$ (0.5\%, which is quite a reasonable accuracy of photo-z for clusters) the significance is $\sim4-5\sigma$,.

Overall, we conclude that the detection of BAO with photometric-quality $z$ is plausible for both AGN and clusters. 
The comparable statistical significance of BAO  detection in both samples is also consistent with our expectation from the preliminary analysis based on effective volumes  (sect. \ref{initial_feeasibility}). 

\subsection{Cosmological parameters}
\label{results_cosmo}

For each configuration of the photo-z quality, we calculate the Fisher matrix $F$ of the cosmological parameters and the information content number called Figure of Merit (FoM),
$$FoM = log\frac{\pi}{\sqrt{det(F^{-1})}}$$
Hence for every pair $\sigma_0$,$f_{\rm fail}$ one obtains a Fisher matrix and the FoM.

\begin{figure*}
    \centering
    \includegraphics[width=0.9\hsize]{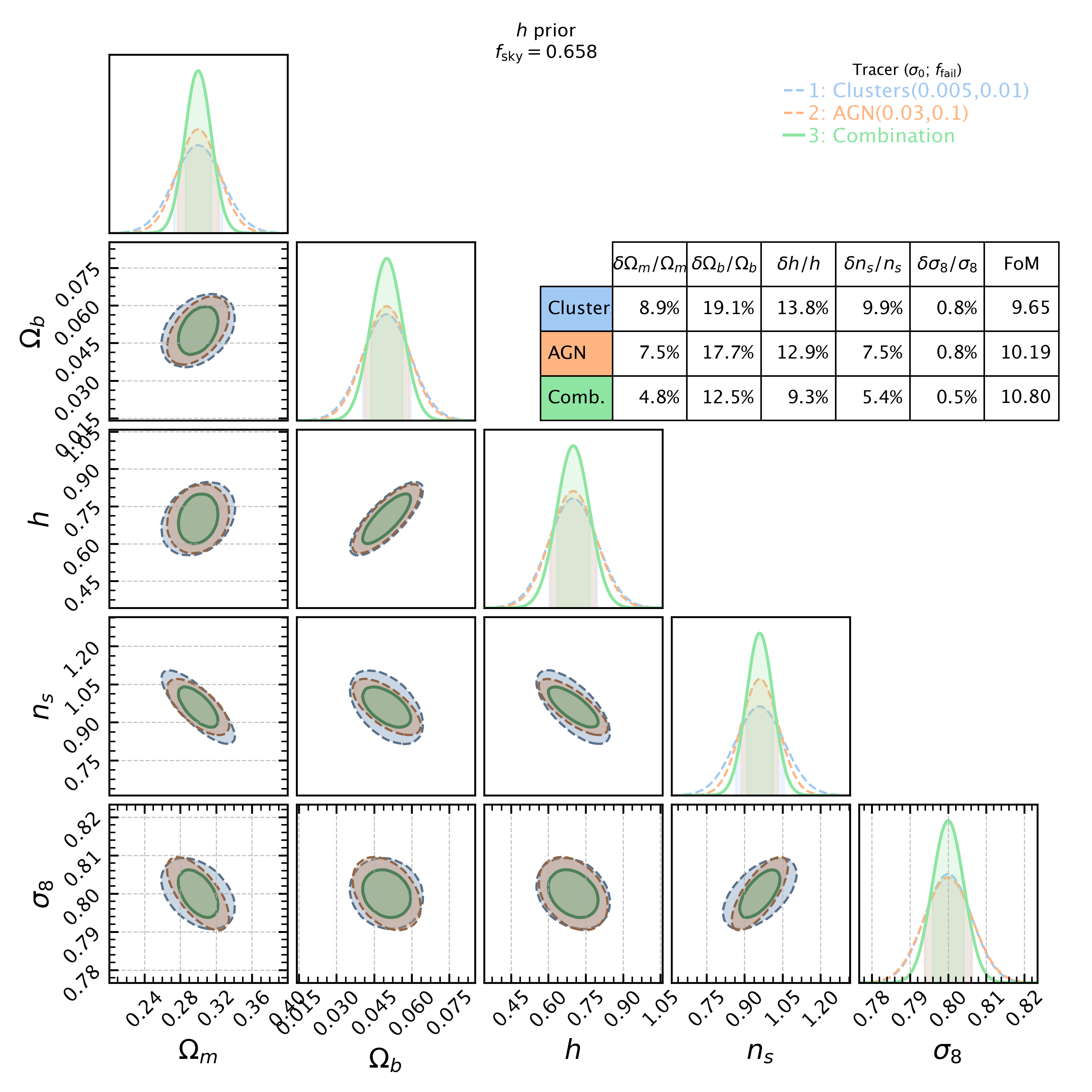}
    \caption{Expected error ellipses of cosmological parameters from the analysis  of  AGN and galaxy clusters in the complete \ero\, all-sky survey. Photometric redshift quality $\sigma_0 = 0.03, 0.005$ and $f_{\rm fail} = 0.1, 0.01$ for AGN and clusters respectively. Prior 0.1 on $h$ is added to the Fisher matrix. The corner plot shows the results derived from the Fisher matrix of clusters (blue), AGN (orange) and combined (green) samples. The table shows the percentage errors on the parameters, and the value for the Figure of Merit.}
    \label{fig:fisher}
\end{figure*}

Before we proceed with the effects of photo-z quality on the forecast, we show one example for the case $\sigma_0 = 0.03, 0.005$ and $f_{\rm fail} = 0.1, 0.01$ for AGN and clusters respectively, see Fig. \ref{fig:fisher}. We add a relatively weak Gaussian prior  ($0.1$) on parameter $h$ for this illustration. The value in the parentheses is the standard deviation of the prior distribution and is indicated for other priors in the rest of the paper. 

We show expected error ellipses from the analysis of both samples separately and in combination (i.e. adding two Fisher matrices). The figure also shows the table with the expected marginalized errors in per cent. Both AGN and cluster samples give comparable errors on all the cosmological parameters.  More importantly, the combination of AGN with clusters  increases the FoM of the former by $0.6$. This is due to the partial alleviation of the parameter degeneracy. 

Parameter $\sigma_8$ is well-constrained because  the error on the amplitude of the signal directly yields the constraints on the amplitude of fluctuations $\sigma_8$. 
Parameters $\Omega_c$ and $n_s$ have accuracy  better than $\sim5\%$ because they control the overall shape of the matter power spectrum and the prior added on $h$ which lifted the degeneracy with $n_s$ and $\Omega_c$. $\Omega_b$ is not well-constrained since the significance of baryonic effects on the angular power spectrum is limited.

\begin{figure*}
    \centering
    \includegraphics[width=0.9\hsize]{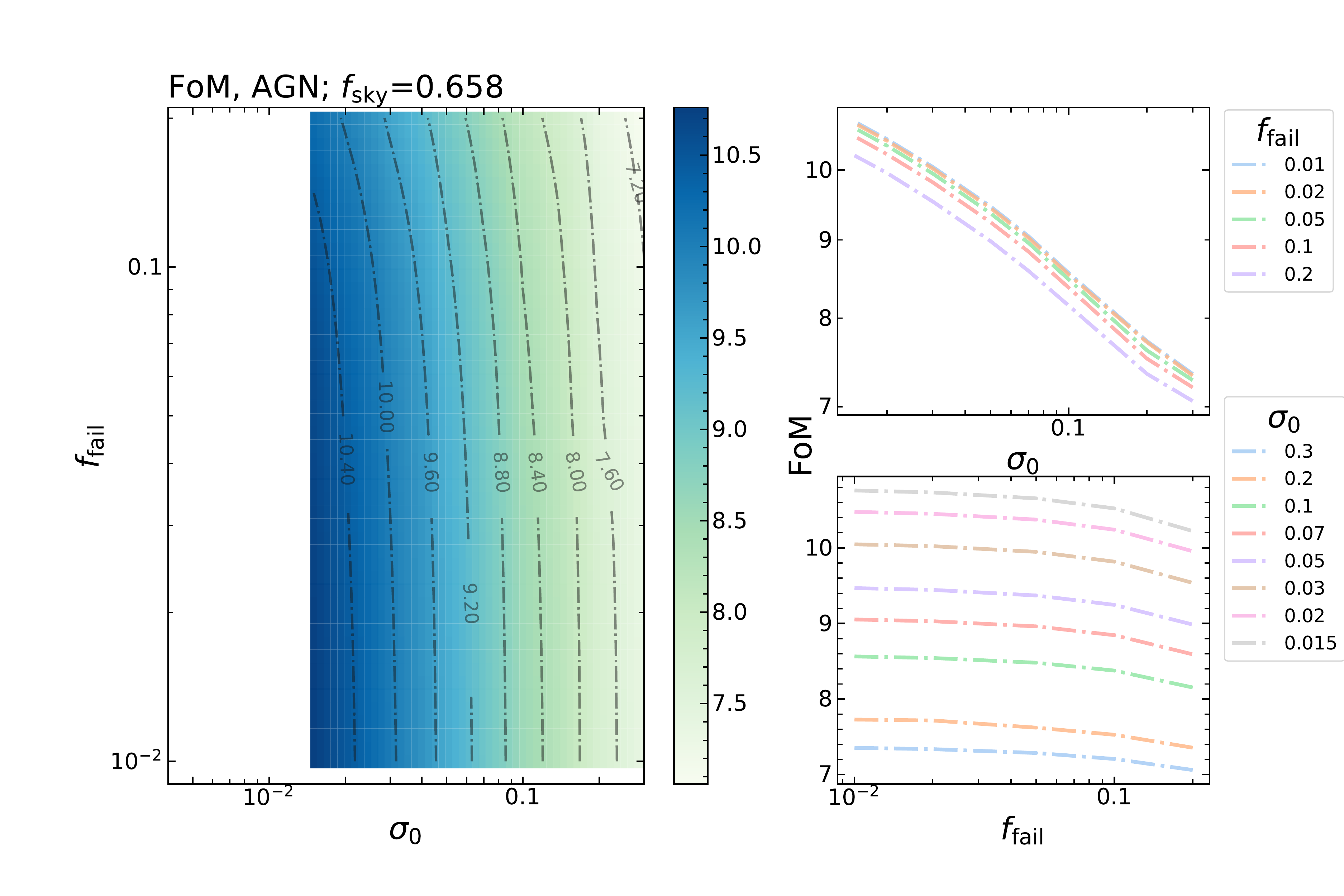}
    \includegraphics[width=0.9\hsize]{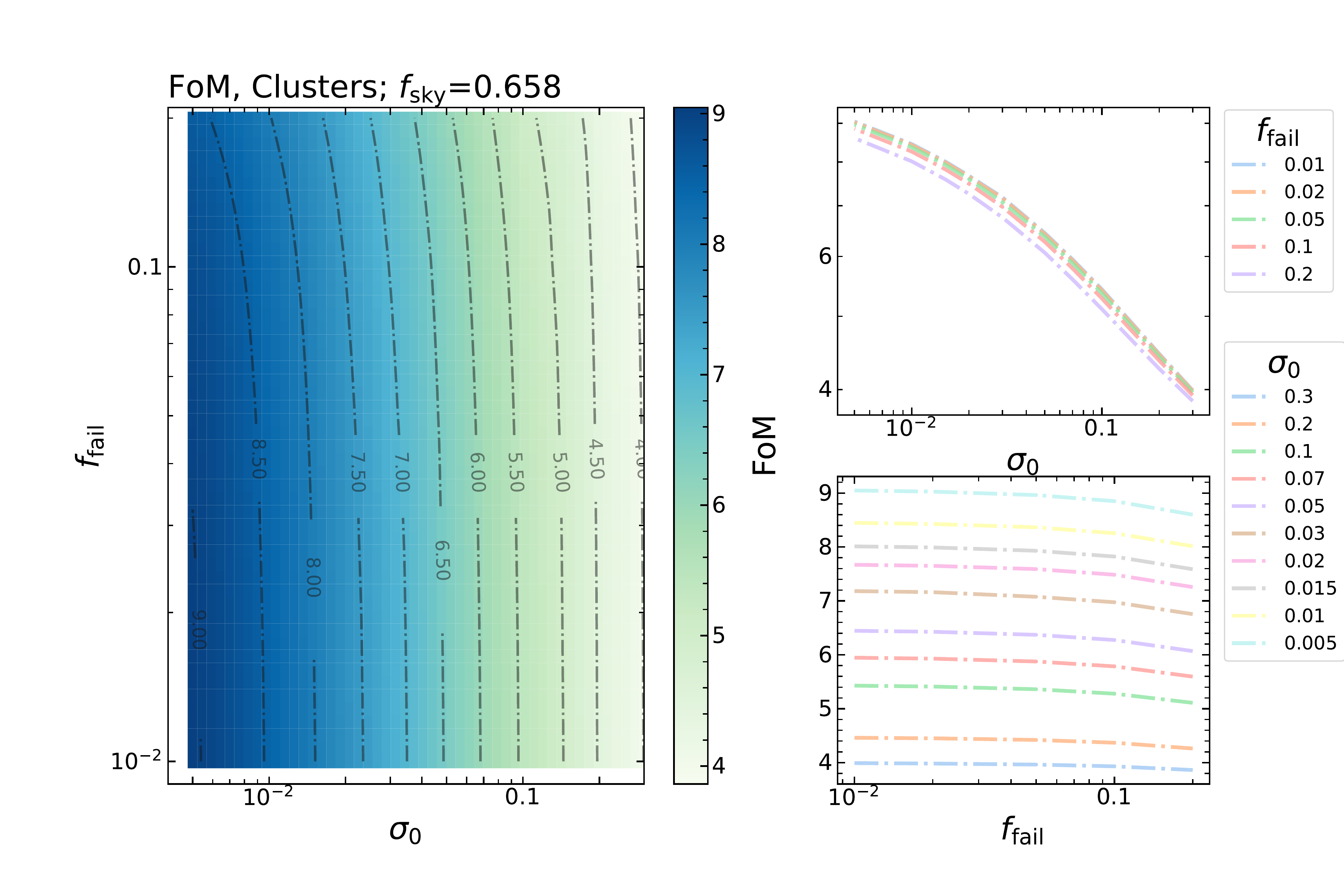}
    \caption{Quality of cosmological constraints (Figure of Merit) in the sample of AGN (top sub-figure) and clusters (bottom). On the left panel, the contours of quality are shown as a function of photo-z parameters  $\sigma_0$ and $f_{\rm fail}$. On the right panels, the cross-sections with constant $f_{\rm fail}$ (top) and $\sigma_0$ (bottom) are shown.}
    \label{fig:cosmo}
\end{figure*}

We made such a forecast for each combination of photo-z precision parameters and condense the result in Fig. \ref{fig:cosmo} using  FoM as the proxy for the quality of the forecast. 
We assumed no priors for this calculation.

As with BAO, photo-z scatter  $\sigma_0$ affects the analysis quality dramatically, while $f_{\rm fail}$ has a less profound influence on the result. For instance, decrease of $\sigma_0$ from 0.1 to 0.03 increases the FoM from $\sim8.5$ to $\sim10$ for AGN with $f_{\rm fail}=0.01$. For clusters $FoM  \sim 6.5$ with $\sigma_0=0.05$ and $FoM\sim9$ if $\sigma_0=0.005$.   We conclude that the quality of  cosmological analysis pivots on the $\sigma_0$ parameter. 

Table \ref{tab:cosmo} shows the quality of the forecast for several cases of photometric redshift parameters. For a 'conservative' case we choose  the following parameters: $\sigma_0 = 0.01, 0.05$ and $f_{\rm fail} = 0.02, 0.1$ for clusters and AGN respectively. For an 'optimistic' case we choose $\sigma_0 = 0.005, 0.03$ and $f_{\rm fail} = 0.01, 0.1$. We show options without prior information, with a wide  prior on the reduced Hubble constant $h$ (0.1) and, in particular,  the prior knowledge of parameters $h$ (0.0054) and $n_s$ (0.0042) derived from the Planck CMB experiment data (CMB+lensing, see table 1 in \citet{PlanckCollaboration2020}, last column). This demonstrates that the combination of the LSS probed by \ero\, and the independent data from other experiments may significantly enhance the resulting constraints. 

\begin{table*}
    \centering

\begin{tabular}{c|c|c|ccccc|c}
\hline
        &                                                 &             & $\Omega_m$ & $\Omega_b$ & $h$    & $n_s$  & $\sigma_8$ & FoM   \\
Photo-z ($\sigma_0$, $f_{\rm fail}$) & Priors                                          & Tracer      &            &            &        &        &            &       \\ \hline
\multirow{9}{*}{Clusters(0.01,0.02); AGN(0.05,0.1)}  & \multirow{3}{*}{no prior} & Clusters & 15.9\% &     81.4\% &  72.4\% &  36.0\% &      1.4\% &   8.43 \\
        &                                                 & AGN         & 10.3\%     & 45.9\%     & 40.4\% & 18.6\% & 1.2\%      & 9.25  \\
        &                                                 & combination & 7.0\% &     38.0\% &  33.8\% &  15.5\% &      0.8\% &   9.77 \\ \cline{2-9} 
        & \multirow{3}{*}{$h$ prior (0.1)}                  & Clusters    & 11.7\% &     22.2\% &  14.0\% &  11.5\% &      1.1\% &   9.14 \\
        &                                                 & AGN         & 9.8\%      & 20.4\%     & 13.5\% & 8.6\%  & 1.0\%      & 9.72  \\
        &                                                 & combination &  6.3\% &     14.5\% &   9.7\% &   6.3\% &      0.7\% &  10.31 \\ \cline{2-9} 
        & \multirow{3}{*}{$h$, $n_s$ Planck prior} & Clusters &      5.4\% &     15.8\% &   0.8\% &   0.4\% &      0.9\% &  11.73 \\
        &                                                 & AGN &      5.1\% &     14.2\% &   0.8\% &   0.4\% &      0.6\% &  12.13 \\
        &                                                 & combination &     3.6\% &     10.0\% &   0.5\% &   0.3\% &      0.5\% &  12.74 \\\hline
\multirow{9}{*}{Clusters(0.005,0.01); AGN(0.03,0.1)} & \multirow{3}{*}{no prior} & Clusters & 13.1\% &     61.5\% &  54.8\% &  28.3\% &      1.0\% &   9.05 \\
        &                                                 & AGN         & 8.1\%      & 35.3\%     & 30.6\% & 14.5\% & 1.0\%      & 9.82  \\
        &                                                 & combination & 5.3\% &     27.6\% &  24.5\% &  11.3\% &      0.6\% &  10.38 \\ \cline{2-9} 
        & \multirow{3}{*}{$h$ prior (0.1)}                  & Clusters    & 8.9\% &     19.1\% &  13.8\% &   9.9\% &      0.8\% &   9.65 \\
        &                                                 & AGN         & 7.5\%      & 17.7\%     & 12.9\% & 7.5\%  & 0.8\%      & 10.19 \\
        &                                                 & combination & 4.8\% &     12.5\% &   9.3\% &   5.4\% &      0.5\% &  10.80 \\ \cline{2-9} 
        & \multirow{3}{*}{$h$, $n_s$ Planck prior} & Clusters &       3.9\% &     11.3\% &   0.8\% &   0.4\% &      0.6\% &  12.11 \\
        &                                                 & AGN &      3.9\% &     10.5\% &   0.8\% &   0.4\% &      0.5\% &  12.47 \\
        &                                                 & combination &      2.6\% &      7.3\% &   0.5\% &   0.3\% &      0.4\% &  13.10 \\ \hline
\end{tabular}

    \caption{SRG/\ero\, forecasts. Fisher marginalized errors for cosmological parameters (in \% of the fiducial value) and Figure of Merit values for two cases of photometric redshift quality and different priors added to the Fisher information. Results are shown for AGN and clusters of galaxies separately and in combination. See text for details.}
    \label{tab:cosmo}
\end{table*}

\section{Discussion}
\label{discusion}
\subsection{Comparison with dedicated cosmological surveys}
\subsubsection{BAO}
\label{discuss-bao}
In sect. \ref{results_bao} we predict that assuming realistic accuracy of photo-z determination -- $\sigma_0=0.03$ and 0.005 for AGN and clusters respectively, baryonic oscillations would be significantly detected in the  SRG/\ero\, all-sky survey in the distribution   AGN (redshift range of $0.5<z<2.5$, about 2 million sources) and clusters of galaxies (redshift range $0.1<z<0.8$, about 90000 sources). Those figures are to be compared with the progress  made so far in  detecting  the baryonic features at other wavelengths and using other LSS tracers.

BAO peaks in galaxy distribution were  detected for the first time in SDSS data \citep{Eisenstein2005,Hutsi2006} and have been routinely observed ever since \citep{Ross2015, Alam2021}. 

There are many detections of the BAO with spectroscopic quality information at different redshifts. SDSS galaxies observed in the Baryon Oscillation Spectroscopic Survey (BOSS\footnote{\url{https://www.sdss.org/surveys/boss/}}) provide BAO detection in the redshift range $0.2<z<0.75$ from the sample of 1.2 million galaxies \citep{Alam2017}.
Extended BOSS program (eBOSS\footnote{\url{https://www.sdss.org/surveys/eboss/}}) pushed the redshift range further, $0.6<z<1.1$, with $\sim400$k luminous red galaxies and $\sim200$k emission line galaxies \citep{Gil-Marin2020, Raichoor2021, Bautista2021, deMattia2021}. SDSS observed the BAO in the distribution of quasar ($0.8<z<2.2$ with $\sim350$k objects) and Ly$-\alpha$ systems ($210$k quasars at $z>2.1$), see \citet[][]{Neveux2020, duMasdesBourboux2020, Hou2021}.

Photometric data samples include the  BAO detection in SDSS photometric catalogues ($0.2<z<0.6$, $\sim600$k galaxies) \citep{Padmanabhan2007, Crocce2011, Seo2012}. Recent results from Dark Energy Survey provide evidence for baryonic wiggles in the distribution of $\sim7$ million galaxies with photo-z quality $\sigma_0=0.03$ in redshift range $0.6<z<1.1$ \citep{DESCollaboration2021b}.

For clusters, BAO were tentatively ($\sim2\sigma$) detected in SDSS photometric cluster catalogue with some $\sim10000$ objects at $z<0.3$ \citep{Estrada2009, Hutsi2010} for the first time. The significance of the detection has increased with the advent of better data,  $\sim3.7\sigma$ detection with $\sim80$k  SDSS clusters at $z<0.5$ with spectroscopic redshifts in the work of \citet{Hong2016}, see also \citet[][]{Moresco2021}.

Considering all of the above it is clear that the SRG/\ero\, sample of AGN and clusters  would provide sufficiently precise measurements of BAO (and the corresponding sound horizon scales) to  expand and complement current experiments, and to provide a cross-check of clustering measurements. It may not be so efficient as SDSS or DES galaxies at a low redshift regime ($z<0.6-0.7$), but due to the satisfactory density of high-redshift objects it may compete with BAO findings in eBOSS emission line galaxies and quasars distribution (see also fig. 9 in \citet{Kolodzig2013b}). 

For clusters of galaxies, the SRG/\ero\, sample of X-ray-selected clusters would provide clear detection of BAO even with photometric quality redshifts with significance comparable to or exceeding  the current statistical power of spectroscopic sample of clusters from SDSS. 

\subsubsection{Cosmological parameters}
\label{discuss-cosmo}
In Section \ref{results_cosmo} we forecast the precision of determination of cosmological parameters from the angular clustering of AGN and clusters in the all-sky \ero\, survey. We show that under realistic assumptions about  the accuracy of the photo-z one can achieve $\sim1-20\%$ marginalized errors on the cosmological parameters. The errors can be further reduced by a factor of $\sim 2-3$ if using  prior information from other experiments. 

 In Table \ref{tab:cosmo-comparison} we compare SRG/\ero\, forecasts with other cosmological probes.  In addition to the cosmological parameters discussed earlier, to ease comparison with other LSS surveys, we calculate uncertainty on the parameter of clustering amplitude $S_8=\sigma_8(\Omega_{m}/0.3)^{0.5}$. For comparison, in the two top lines of the table,  we repeat \ero\,     combined AGN and clusters of galaxies forecast from the second half of Table \ref{tab:cosmo}. We show both forecasts made without and with the Planck priors $h$ (0.0054) and $n_s$ (0.0042) \citep{PlanckCollaboration2020}. 
The accuracy of cosmological parameters determination from the Planck data is shown in the third line of the table. Notably, the combined \ero\, sample in combination with Planck priors leads to improvement in the precision of determination for $h$, $n_s$ and $\sigma_8$ compared to Planck-only data. 

Some of the LSS experiments use different LSS   tracers \citep{Alam2021}. For example, \citet{Loureiro2019} used spectroscopic SDSS BOSS data of $\sim1.3$ million galaxies  ($0.15<z<0.8$). The results (their table 4) are listed in the 4th line in Table \ref{tab:cosmo-comparison}. X-ray samples, used without Planck priors, would have tighter errors on the parameter $\Omega_m$ and $S_8$, approximately similar uncertainty on $\Omega_b$ and larger error margins for $h$ and $n_s$. 

The ongoing Dark Energy Survey (DES) measures the shapes and positions of millions of galaxies. DES uses cosmic shear and clustering of galaxies as the effective cosmological probe of the matter fluctuation amplitude and density.  We show their results (Y3, table II in \citet{DESCollaboration2021a}) for flat $\Lambda$CDM from the DES data alone  and with a combination of BAO and RSD measurements, Supernovae and CMB data. We see that SRG data alone provide better constraints on the matter density and performs worse for $\sigma_8$ (we note in our forecast we do not fix the other three parameters of the model). If compared with the combination of the DES with other probes, it is seen that the combination of \ero\, data and Planck prior for two parameters would perform comparably, having similar errors on $h$, $n_s$, larger errors on $\Omega_m,\Omega_b, S_8$ and slightly smaller error on $\sigma_8$.  We also show the forecast results \citep{EuclidCollaboration2020} for \textit{Euclid} probe for the case of photometric galaxy clustering and weak lensing (their table 9, optimistic settings for GC$_{\text{ph}}$+WL). As it should have been expected, Euclid will outperform \ero\, constraints without priors.

\begin{table*}[]
    \centering
    \large
    
\begin{tabular}{c||cccccc}
Experiment                                                                           & $\Delta\Omega_m$ & $\Delta\Omega_b$ & $\Delta h$   & $\Delta n_s$ & $\Delta\sigma_8$ & $\Delta S_8$ \\ \hline\hline
SRG/eRosita$^1$                                                                          &  $\pm$0.016      &  $\pm$0.014      &  $\pm$0.172 &  $\pm$0.108 &  $\pm$0.005 & $\pm$0.019     \\ \hline
\\
\begin{tabular}[c]{@{}c@{}}SRG/eRosita$^1$\\ $h$, $n_s$ Planck prior\end{tabular} & $\pm$0.008      & $\pm$0.004     & $\pm$0.004 & $\pm$0.003 & $\pm$0.003 & $\pm$0.012  \\ \hline \\
Planck (CMB+lensing)$^2$ & $\pm$0.0074 & $\pm$0.0007 & $\pm$0.0054 & $\pm$0.0042 & $\pm$0.0061 &  $\pm$0.013 \\ \hline \\
SDSS BOSS$^3$ & $^{+0.034}_{-0.033}$ & $^{+0.010}_{-0.009}$& $_{-0.069}^{+0.088}$ & $^{+0.064}_{-0.045}$ & -  & $_{-0.064}^{+0.072}$ \\\hline \\ \\
DES 3$\times$2pt$^4$ & $^{+0.032}_{-0.031}$ & - & - & - & $^{+0.039}_{-0.049}$ & $_{-0.017}^{+0.017}$\\ \hline \\ \\
\begin{tabular}[c]{@{}c@{}}DES 3$\times$2pt\\ +BAO+RSD+SNIa+CMB$^4$\end{tabular}  & $^{+0.004}_{-0.005}$ & $_{-0.0004}^{+0.0005}$ & $_{-0.003}^{+0.004}$ & $^{+0.004}_{-0.003}$ & $_{-0.005}^{+0.008}$ & $_{-0.008}^{+0.008}$ \\ \hline \\
SRG/eRosita & $\pm$0.0031 & $\pm$0.0492 & $\pm$0.364 & $\pm$0.143 & $\pm$0.003 & - \\ 
cluster mass function$^5$\\ \hline \\
\textit{Euclid}$^6$ & $\pm 0.0038$ & $\pm0.046$& $\pm0.020$ & $\pm0.0037$ & $\pm0.0017$  & - \\\hline \\ \\
\end{tabular}

    \caption{ Comparison of SRG/\ero~ forecasts with other cosmological probes. Forecast Fisher marginalized errors for cosmological parameters in the flat $\Lambda$CDM model. SRG/\ero~ results are for the combined data of  AGN and clusters of galaxies, with and without Planck priors.\\ References:  $^1$: this work; $^2$: \citet{PlanckCollaboration2020}; $^3$: \citet{Loureiro2019}; $^4$: \citet{DESCollaboration2021a}; $^5$: \citet{Pillepich2012} ; $^6$: \citet{EuclidCollaboration2020}}
    \label{tab:cosmo-comparison}
\end{table*}

 The above comparison demonstrates  that samples of AGN and clusters of galaxies from SRG/\ero\, all-sky survey combined with the currently available photo-z estimates provide a sufficiently powerful cosmological LSS probe. They  compete in statistical power  with those derived from the dedicated cosmological large-scale structure surveys in the optical band. The statistical power of the \ero\, X-ray-selected samples will increase further if and when the accuracy of redshift determination achieves spectroscopic quality. 
This appears to be  feasible, in principle, for clusters of galaxies which require acquiring  about $\sim10^5$ optical spectra. For AGN, where of the order of  $\sim10^6$ new optical spectra need to be obtained, this may be a more difficult task for a more distant future.

The calculations for cluster mass function were performed in \citet{Pillepich2012}.
We use the results for the fixed M-L relation for clusters of galaxies. They took the accuracy of photo-z into account by setting the redshift bin width at $0.05(1+z)$, assumed sky coverage  of $f_{\rm sky}=0.658$ and also used the Fisher matrix formalism. Comparing the last row of Table \ref{tab:cosmo-comparison} with the first two rows and with Table \ref{tab:cosmo} one can conclude  that the cluster mass function is an equally important cosmological tool. When both are used  without any priors, the mass function-based measurement would be about $\sim 5$ times more accurate in measuring $\Omega_m$ than the LSS-based one (which is to be expected).  However, all other cosmological parameters are determined more accurately, by a factor of $\sim 1.5-3.5$ in the LSS-based measurement (Clusters+AGN). Expressed in terms of the Figure of Merit, the cluster mass function is expected to achieve $FoM\approx 10.1$  which should be compared with $FoM\approx 9.8-10.4$ for LSS-based measurement (Table \ref{tab:cosmo}). The power of the cluster mass function method increases significantly when combined with the Planck priors. 

Finally, it should be noted that the cluster mass function is prone to systematic errors in mass determination, whereas the LSS-based measurement does not include systematic uncertainties of comparable amplitude (at least for AGN) and therefore should be more robust. As discussed in \citet{Pillepich2012}, the Figure of Merit of cosmological analysis degrades dramatically, when the uncertainty in the   M-L relation is included.

\subsection{Dependence on the parameters of the all-sky survey.}
\label{discuss-survey}

SRG satellite performs a full scan of the sky in 6 months \citep{Sunyaev2021} and it is planned to conduct 8 all-sky surveys in the course of the mission. After the 4.4 already completed  surveys, \ero\,  achieved the record all-sky sensitivity of about $\sim 15$ times better that reached  in the previous all-sky survey performed by  ROSAT satellite in 1990 \citep{Truemper1982, Voges1999}.
It is therefore interesting to access the quality of cosmological measurements which can be achieved before the completion of the full survey.  

To this end, we rerun our calculations  changing the limiting flux of the survey to the values corresponding to  2 years of the survey (named eRASS4) and to 3 years (eRASS6). For point sources these values are $1.45\times10^{-14}$ (eRASS4) and $1.25\times10^{-14}$  (eRASS6) \ergpspcm, see \citep{Kolodzig2013a}. Changing the limiting flux affects the redshift distribution of tracer objects and, more importantly, their number density within redshifts of interest. For eRASS4,6,8 the number of sources is 1\,051\,773, 1\,346\,620  and 1\,914\,915  respectively ($0.5<z<2.5$). 
In Table \ref{tab:cosmo-surv-params} the forecast (with wide prior in $h$ and photo-z parameters $\sigma_0=0.03$ and $f_{\rm fail}=0.1$) is shown for these limiting fluxes. FoM values between eRASS4 and eRASS8 differ by $0.8$,  and the errors on the cosmological parameters increase by a  factor of $\sim 1.5-2$.

We also investigated the impact of the survey area on the forecast. We start by considering the case when only data from half of the extragalactic sky is used in calculations.
We neglect the effects of the mode coupling in angular power spectra of half-sky masked data and simply change the sky area surveyed by a factor of 2. In addition,  we present a forecast for the survey with the area of $9000$ deg$^2$ which  approximately corresponds to the footprint of the SDSS survey where the model of photo-z would be better calibrated (at least for the AGN sample).

The scaling of Fisher matrices with $f_{\rm sky}$ is trivial, $F_{i,j}\propto f_{\rm sky}$, hence the marginalized errors scale as the inverse square root of the sky area.  We illustrate in Table \ref{tab:cosmo-surv-params} the results for the fiducial eRASS8 AGN setup and show  errors on cosmological parameters obtained for different sky areas used in the analysis. It can be seen that for the one half of the extragalactic sky $f_{\rm sky} = \frac{0.658}{2}$ the results differ from the all-extragalactic case by a factor of $\sqrt{2}$, and the Figure of Merit differ by $0.75$, whereas for $9$k deg$^2$ case the errors are larger by approximately a factor of two, and FoM value smaller by $1.2$.

\begin{table*}[]
    \centering
    \large
    \begin{tabular}{l|cccccc}

{} & $\Omega_m$ & $\Omega_b$ &     $h$ &   $n_s$ & $\sigma_8$ &    FoM \\ \hline

eRASS4               &     11.4\% &     22.4\% &  13.7\% &   9.7\% &      1.1\% &   9.39 \\
eRASS6              &      9.6\% &     20.1\% &  13.5\% &   8.6\% &      1.0\% &   9.72 \\
eRASS8                &      7.5\% &     17.7\% &  12.9\% &   7.5\% &      0.8\% &  10.19 \\
eRASS8,  half extragal.      &     10.7\% &     25.1\% &  18.3\% &  10.7\% &      1.1\% &   9.44 \\
eRASS8, 9000 deg$^2$ &     13.1\% &     30.7\% &  22.4\% &  13.1\% &      1.4\% &   9.00

\end{tabular}

    \caption{SRG/\ero\, forecasts for different survey parameters. Forecast Fisher marginalized errors (in \% of fiducial value) on AGN clustering depending on the survey parameters (survey depth and solid angle). eRASS4,6 and 8 correspond to 2,3 and 4 years of the all-sky survey. Photo-z accuracy parameters are $\sigma_0=0.03$ and $f_{\rm fail}=0.1$. See Section \ref{discuss-survey} for further details.}
    \label{tab:cosmo-surv-params}
\end{table*}

\section{Conclusions}
\label{conclusions}

We  investigated the potential of X-ray-selected samples  of AGN and clusters of galaxies (to be) detected in the SRG/\ero\, all-sky survey  to serve as a cosmological probe.  We focused  on the ability to detect Baryon acoustic oscillations (BAO) and to constrain cosmological parameters under the assumption of the availability of photometric redshifts  of realistically achievable quality.
Our main results are obtained in sect. \ref{results_bao} and \ref{results_cosmo}.

Using the model of \citet{Hutsi2014b} of  photometric redshift scatter we show that for both BAO and cosmological forecast the redshift scatter parameter $\sigma_0$ has more influence on the quality of cosmological constraints than the fraction of catastrophic errors $f_{\rm fail}$ (Fig.\ref{fig:bao},\ref{fig:cosmo})

We demonstrate that under reasonable assumptions regarding the quality of photo-z  ($\sigma_0=0.03$ and 0.005 for AGN and clusters of galaxies respectively) it is possible to detect BAO with significance $\sim5-6\sigma$ and $\sim4-5\sigma$ in the distribution of AGN and clusters respectively. This is comparable with the BAO detections in large-scale structure surveys  for galaxies and clusters (sect. \ref{discuss-bao}).

Fisher matrix analysis of angular power spectra under the same assumptions yields: (i) a joint analysis of AGN and cluster data alleviates some of the degeneracies and reduces errors on the cosmological parameters by a factor of $\sim1.5$ (Fig. \ref{fig:fisher}), (ii) solely X-ray data constrain cosmological parameters with the accuracy in the  $\sim5-25\%$ range without priors and in the  $\sim0.5-10\%$ range with Planck priors (Table \ref{tab:cosmo}), (iii) X-ray-selected samples of SRG/\ero\, AGN and clusters of galaxies used solely or in combination with other data is a powerful cosmological probe which is quite competitive with the dedicated cosmological surveys like SDSS or DES (sect. \ref{discuss-cosmo} and Table \ref{tab:cosmo-comparison}).

Comparing with  the results of \citet{Pillepich2012} we conclude that cosmological measurements based on the mass function of clusters of galaxies are expected to provide about $\sim 5$ times more accurate  measurement of $\Omega_m$ than the LSS-based one. However, all other cosmological parameters are determined by a factor of $\sim 1.5-3.5$ more accurately  in the clustering-based measurement of AGN and clusters of galaxies. Both methods give comparable values of the Figure of Merit (Table \ref{tab:cosmo}, \ref{tab:cosmo-comparison}).

We investigate the dependence of our forecasts on the survey parameters -- its area and depth  (sect. \ref{discuss-survey}, Table \ref{tab:cosmo-surv-params}). We demonstrate that even with incomplete sky coverage or limited exposure SRG/\ero\, all-sky survey data still produce competitive results.

\begin{acknowledgements}
We are grateful to the referee for useful comments which improved the presentation of findings. SDB acknowledges support from and participation in the International Max-Planck Research School (IMPRS) on Astrophysics at the Ludwig-Maximilians University of Munich (LMU). SDB acknowledges partial support by the subsidy 671– 2020–0052 allocated to Kazan Federal University for assignments in scientific activities. MG and RS acknowledge the partial support of this research by grant 21-12-00343 from the Russian Science Foundation.
Software: CAMB\citep{Lewis2011}, CCL\citep{Chisari2019},  NumPy \citep{Harris2020}, Matplotlib \citep{Hunter2007},  SciPy \citep{2020SciPy-NMeth}, Pandas\citep{reback2020pandas}, ChainConsumer \footnote{\url{https://samreay.github.io/ChainConsumer/index.html}}, COBAYA \citep{Torrado2021}, AstroPy \citep{astropy:2018}, HEALPix \footnote{\url{https://sourceforge.net/projects/healpix/}} \citep{Gorski2005}, HEALPy \citep{Zonca2019}. 
The code used to produce the results of the paper would be available shortly after the publication \footnote{\url{https://github.com/SergeiDBykov/forecast_clustering}}. 
 Data: no data is used for this paper.
\end{acknowledgements}

\bibliographystyle{aa} 
\bibliography{literature.bib} 

\begin{appendix} 
\section{Fisher formalism and MCMC method}
\label{mcmc}
There are conditions to be met for the Fisher analysis to yield realistic constraints on the parameters of the model. One such condition is that the posterior probability distribution is a multi-dimensional Gaussian distribution. This is clearly not the case for cosmological models since there are non-linear degeneracies in the parameters. However, one might hope that the resulting errors would be small so that the linear approximation holds and the Fisher matrices method indeed produces a sound forecast.

\begin{figure*}
    \centering
    \includegraphics[width=0.9\hsize]{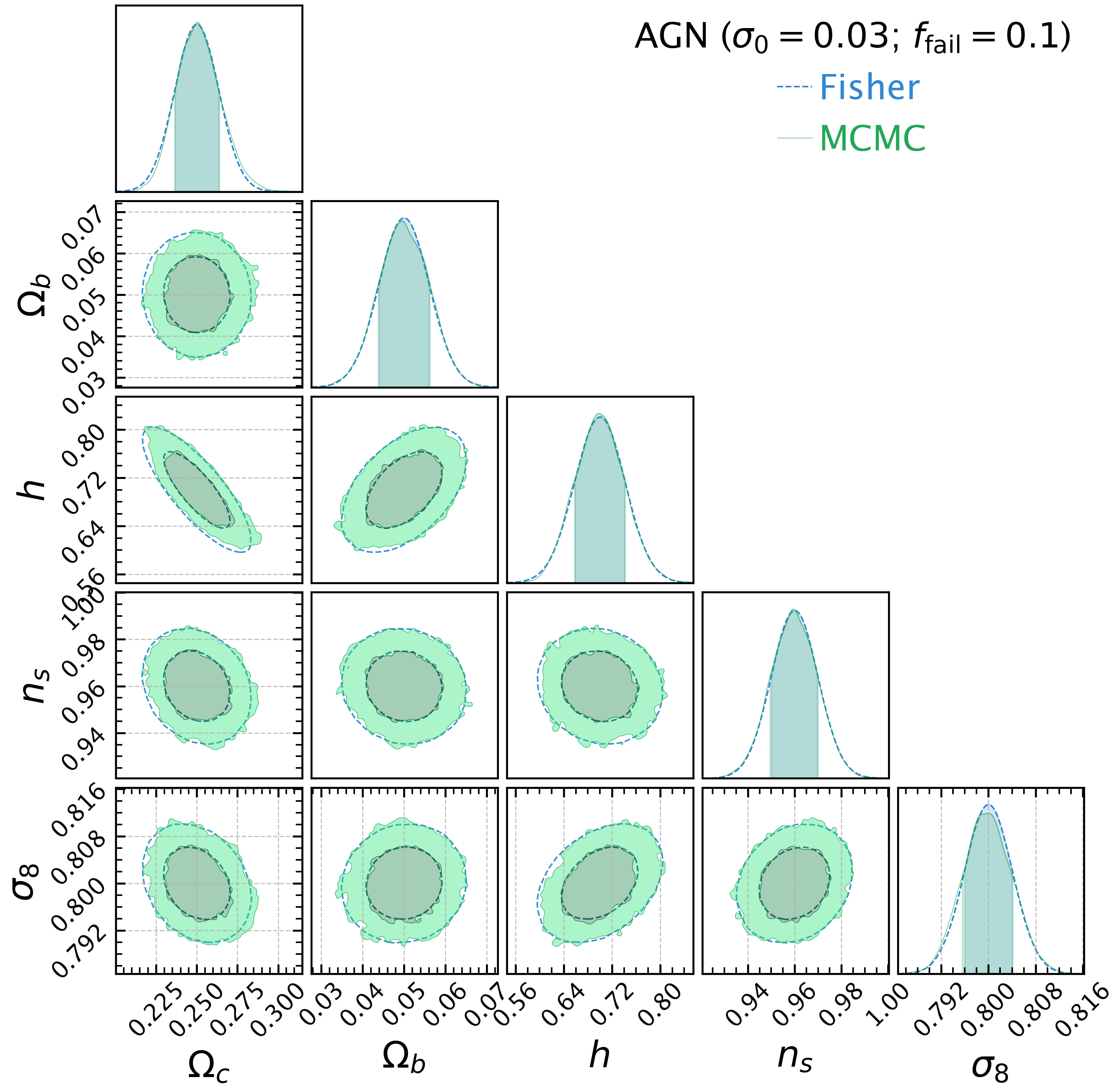}
    \caption{Error ellipses of cosmological parameters estimation from the angular power spectra for the case of AGN ($\sigma_0=0.03$, $f_{\rm fail} = 0.1$) with priors. Blue dashed ellipses are the result of the Fisher forecast, whilst the shaded regions show the result of MCMC sampling done in COBAYA. Darker areas correspond to the 68\% probability contour and the lighter areas to 95\%. On the diagonal of the plot grid, the marginalized histograms of the corresponding parameter are shown. Fisher formalism provides a forecast in good agreement with the MCMC results.}
    \label{fig:mcmc}
\end{figure*}

To test the method, we made a Fisher forecast for the case of AGN (eRASS8, $f_{\rm sky} = 0.658$) with follow-up quality $\sigma_0=0.03$ and $f_{\rm fail} = 0.1$ and priors $0.05$ on $h$ and $0.01$ on $n_s$. Then we used COBAYA\footnote{\url{https://cobaya.readthedocs.io}} code for Bayesian analysis in Cosmology \citep{Torrado2019, Torrado2021} and their implementation of MCMC Metropolis sampler \citep{Lewis2002} to probe the posterior. We assumed Gaussian likelihood 
with appropriate priors and make 'data' from our data vector, loaded a covariance matrix, and then run chains until the convergence.  In Fig. \ref{fig:mcmc} we visualise the probability contours of two methods and in Table \ref{tab:mcmc} we compare the marginalized 68\% credible intervals.  One can see that under our assumptions the Fisher forecast produces contours and errors consistent with the result of the full MCMC treatment of the problem.  Based on the results of this and other similar tests we chose to use Fisher matrix formalism, which is significantly faster computationally than MCMC, as our baseline tool.

\begin{table*}
    \centering
    \caption{MCMC vs. Fisher formalism}
    \label{tab:model_params}
    \begin{tabular}{c|ccccc}
    
		Method & $\Omega_c$ & $\Omega_b$ & $h$ & $n_s$ & $\sigma_8$ \\ 
		\hline
		MCMC &  $0.250^{+0.013}_{-0.014}$ & $\left( 49.9^{+6.3}_{-6.1} \right) \times 10^{-3}$ & $0.696^{+0.045}_{-0.038}$ & $\left( 959.4^{+10.3}_{-10.0} \right) \times 10^{-3}$ & $\left( 800.1^{+4.0}_{-4.5} \right) \times 10^{-3}$ \\ 
		Fisher & $0.250^{+0.013}_{-0.013}$ & $\left( 50.0^{+6.0}_{-6.0} \right) \times 10^{-3}$ & $0.700^{+0.041}_{-0.041}$ & $\left( 960.0^{+9.8}_{-9.8} \right) \times 10^{-3}$ & $\left( 800.0^{+4.0}_{-4.0} \right) \times 10^{-3}$ \\
		
    \end{tabular}
\label{tab:mcmc}
\caption{Table corresponding to Fig. \ref{fig:mcmc}. The values of the mean of the parameters and corresponding marginalized errors are shown for the MCMC method and the Fisher forecast.}
\end{table*}

\end{appendix}

\end{document}